\documentclass[12pt]{iopart}

\usepackage[]{graphicx}
\usepackage{url}
\usepackage[T1]{fontenc}
\usepackage{dcolumn}
\usepackage{sistyle}
\usepackage{multirow}
\usepackage{color}
\usepackage{booktabs}
\usepackage{threeparttable}
\usepackage{appendix}

\begin{document}

\title{Wideband digital phase comparator for high current shunts}

\author{Umberto Pogliano, Bruno Trinchera and Danilo Serazio}

\address{Istituto Nazionale di Ricerca Metrologica (INRIM),
 Str.\ delle Cacce 91, 10135 Torino, Italy}

\eads{\mailto{u.pogliano@inrim.it} and \mailto{b.trinchera@inrim.it}}

\begin{abstract}

A wideband phase comparator for precise measurements of phase difference of high current shunts has been developed at INRIM. The two-input digital phase detector is realized with a precision wideband digitizer connected through a pair of symmetric active guarded transformers to the outputs of the shunts under comparison. Data are first acquired asynchronously, and then transferred from on-board memory to host memory. Because of the large amount of data collected the filtering process and the analysis algorithms are performed outside the acquisition routine. Most of the systematic errors can be compensated   by a proper inversion procedure. 

The system is suitable for comparing shunts in a wide range of currents, from several hundred of milliampere up to \SI{100}{A}, and frequencies ranging between \SI{500}{Hz} and \SI{100}{kHz}. Expanded uncertainty (k=2) less than \SI{50}{\mu rad}, for frequency up to \SI{100}{kHz}, is obtained in the measurement of the phase difference of a group of \SI{10}{A} shunts, provided by some European NMIs, using a digitizer with sampling frequency up to \SI{1}{MHz}. An enhanced version of the phase comparator employs a new digital phase detector with higher sampling frequency and vertical resolution. This permits to decrease the contribution to the uncertainty budget of the phase detector of a factor two from \SI{20}{kHz} to \SI{100}{kHz}. Theories and experiments show that the phase difference between two high precision wideband digitizers, coupled as phase detector, depends on multiple factors derived from both analog and digital imprint of each sampling system.

\end{abstract}




\maketitle

\section{Introduction}

Electric power measurement and modern instrumentation for power quality monitoring, which represent the state-of-the-art in the traceability of power distribution, require the solution of several problems as well as their metrological characterization. Complex electric quantities are computed in these measurements and, even in the simplest case of a single-phase line, at least two correlated quantities (one voltage and one current) must be taken into account.

In digital sampling power-measurement system, the main problems are linked to the precision and the correlation of the digitizer inputs, and to the characterization of the transducers. These transducers should be stable and insensitive to temperature and signal level and, furthermore, they must have low and stable phase error within the bandwidth of interest.

In particular, precision shunts for alternating current measurement are extensively used in metrological laboratories, and some of them have been specifically developed in order to scale-up the ac-dc current transfer standard \cite{Inglis92, Kinard91, Pogliano97, Budovsky05, Funck07, Zachovalova08}. Nevertheless, for their application to power measurement, particularly for wideband application, besides the accurate determination of the impedance, they also need a phase characterization.

In order to improve the traceability of measurement systems for power and power quality a cooperation has been undertaken among some European National Metrology Institutes (NMIs) in the  framework of iMERA Plus project "The Next Generation of Power and Energy Measurements"\footnote{The research presented in this paper is part of the EURAMET joint research project on "Power and Energy" and has received funding from the European Community's Seventh Framework Programme, ERA-NET Plus, under Grant Agreement No. 217257.}. For precise measurements of alternating current this project aims to develop shunts that are applicable to different current ranges, with high stability of the impedance with load current, temperature and time. The two prevalent technologies already available are the so-called cage designs, where a number of resistive elements are connected in parallel, in a cage-like design \cite{Filipsky06, Lind08, Rydler05}, and coaxial foil structures employing resistive metal alloys  arranged in coaxial structures \cite{Voljc09, Garcocz04}. Both methods  aim at minimizing the inductance that, for these shunts, is the  major reason of phase error. In order to evaluate different designs of shunts, in the framework of the project, measuring systems are developed for their characterization. 

Considering the previous experience in  shunt design and development \cite{Pogliano09}, in the construction of precise transconductance amplifier \cite{Pogliano92} and implementation of analog to digital converters (ADCs) for high precision measurement of electrical power \cite{Pogliano01} at INRIM, a task concerning the  determination of the phase in real working conditions was  proposed. For this purpose, a new type of phase comparator has been developed, based on some concepts applied on a previous phase comparator \cite{Budovsky07}. A first edition, suitable for  measurements of phase differences between two shunts, in the  range between \SI{2}{A} to \SI{100}{A}, for frequencies from \SI{500}{Hz} to \SI{100}{kHz}, has been built. This version has been employed to compare existing shunts produced by the laboratories taking  part in the project, in view of an improvement of their design.

An improved edition of the phase comparator employs a faster digitizer board with \SI{16}{bits} of resolution and maximum sampling rate of \SI{15}{MSamples/s} (i.e. resolution scales up when the sampling rate decreases). It is characterised at fixed sampling frequency evaluating the behaviour of phase difference varying the amplitude and frequency of the applied signal. Nevertheless, the noise of the phase detector, within the band of interest (i.e., up to \SI{100}{kHz}), is given in terms of time domain by a two-sample deviation computed by long-term time series of measurements performed applying single tones of exciting signals at fixed amplitude, keeping  the main parameters of the digitizer constant. This phase detector permits to improve the repeatability of the measurements and to extend over the band of the phase comparator.

\section{Description of the phase comparator}

\begin{figure}[h]
    \centering
    \includegraphics[width=3.5 in]{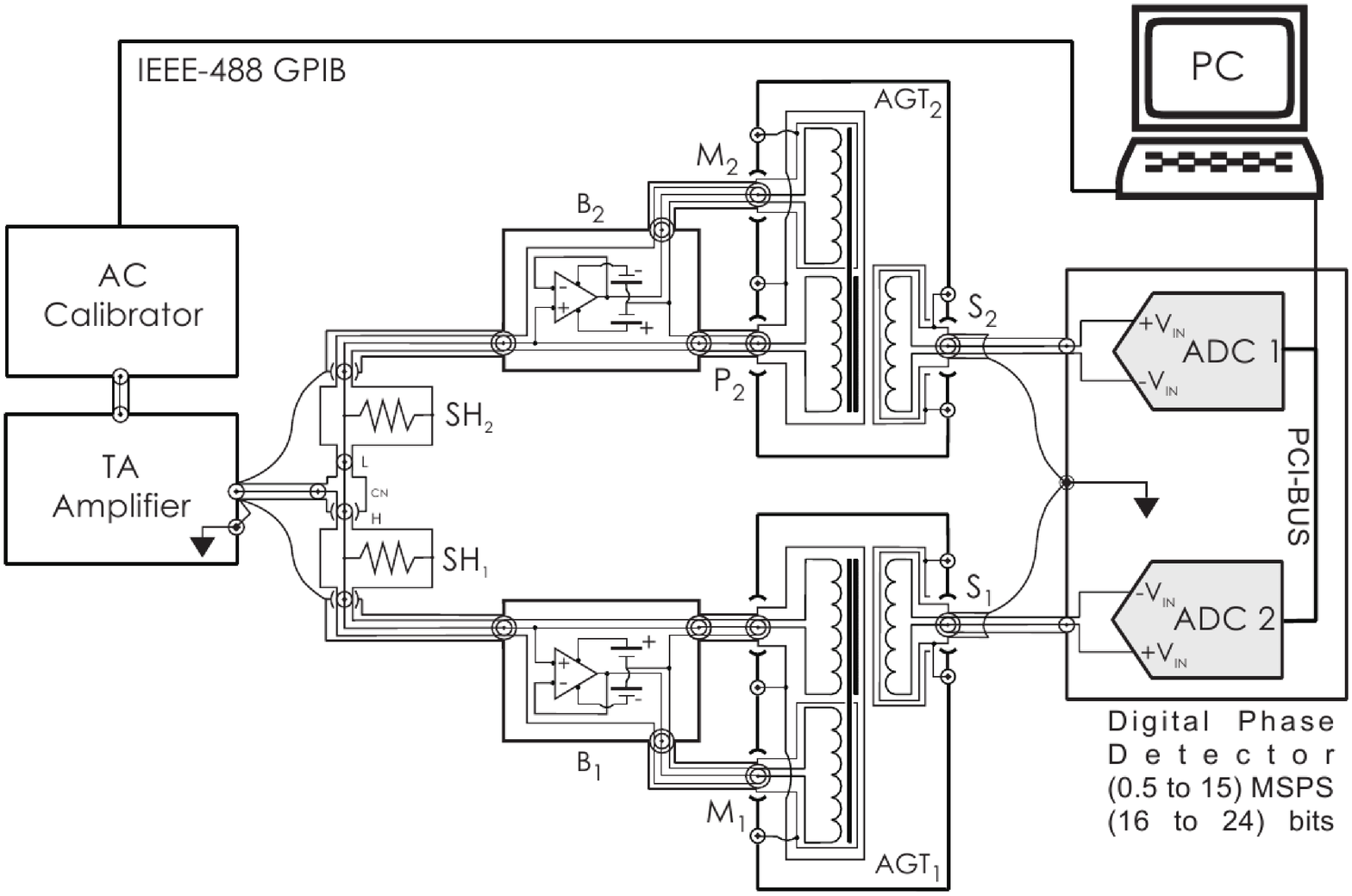}
    \caption{The scheme of the asynchronous wideband digital phase comparator.} 
    \label{fig:SchPhComp}
\end{figure}

The phase comparator has been set up by assembling commercial instruments with electrical and electronic parts specifically designed and built. Figure \ref{fig:SchPhComp} shows the set-up of the sampling system suitable for measuring the phase difference of shunts under comparison. It consists of a system for generating a suitable current flowing through the two shunts, $\mathrm{SH_1}$ and $\mathrm{SH_2}$, which are connected in series. The outputs of the shunts to be compared are then connected through two active guarded transformers, $\mathrm{AGT_1}$ and $\mathrm{AGT_2}$, employed as wideband decoupled precision transmitters, to a two-input digital phase detector.

The systems for current generation and phase detection are controlled by a computer that, in sequence, sets the current amplitude and the frequency programmed, acquires the samples from the digital phase detector, processes the data and evaluates the phase differences (see section \ref{sec:DetPhasDiff} for further details).

\subsection{The system for current generation}
\begin{figure}[h]
    \centering
    \includegraphics[width=3.5 in]{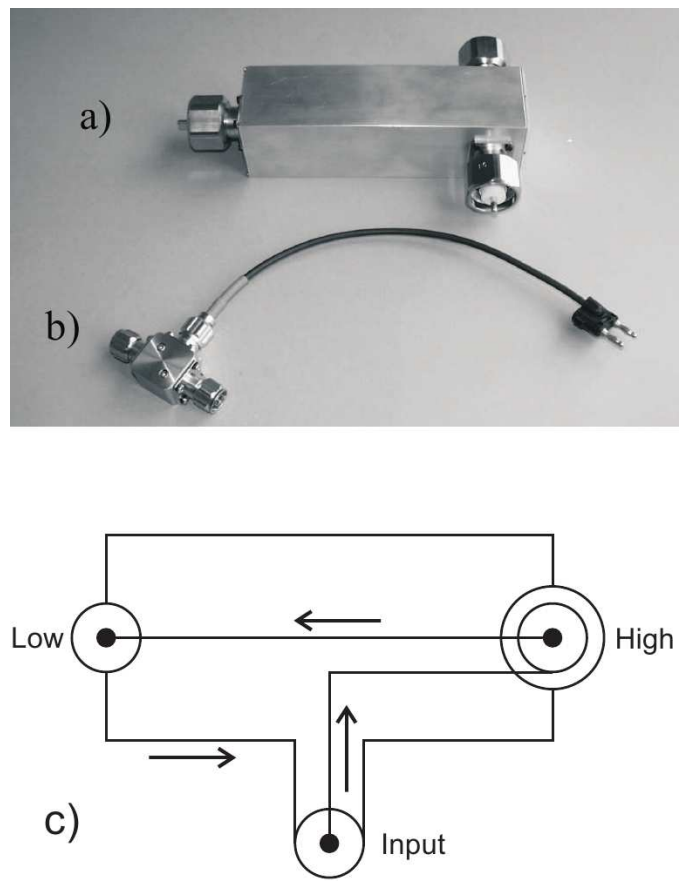}
    \caption{a) Special T-connector for currents from \SI{20}{A} to \SI{100}{A}; b) One type of T-connector for currents lower than \SI{20}{A}; c) The scheme of the internal connections.} 
    \label{fig:CurrentNode}
\end{figure}

The ac current supplying system was assembled by connecting together a calibrator\footnote{Wavetek-Datron 4808 Multifunction Precision Calibrator. Brand names are used for identification purpose. Such use does not imply endorsement by INRIM or assume that the equipment is the best available.} in the ac voltage function and a transconductance amplifier\footnote{Clarke-Hess model 8100.}. The ac current for every specified measurement point is generated at the output of the transconductance amplifier. By means of suitable current nodes (see figure \ref{fig:CurrentNode}(a) for currents from \SI{20}{A} to \SI{100}{A}, figure \ref{fig:CurrentNode}(b) for currents up to \SI{20}{A} and figure \ref{fig:CurrentNode}(c) for internal connections) the alternating current is supplied to the two shunts under comparison which are connected electrically in series. 

The current ranges are established by the transconductance amplifier and they are selectable as shown in table \ref{tab:TAranges}. As the comparator is primarily intended for currents over 5 A, the frequency is limited by the transconductance amplifier to 110 kHz. The maximum current in each range, except for 100 A, is obtained by applying 2 V input (1 V for 100 A) supplied by the calibrator, which is automatically set to the proper voltages and frequencies for the comparison directly by the computer via GPIB (IEEE-488.2) bus.

\begin{table}[ht]
	\caption{Ranges of the transconductance amplifier used in the set-up of the phase comparator as a function of the input voltage applied.}
	\centering
		\begin{tabular}{ccc} \toprule
		\multicolumn{1}{c}{}& \multicolumn{2}{c}{\small{Transconductance Amplifier}} \\
		\cmidrule(r){2-3}
	    \small{AC Calibrator} & \small{Output current} & \small{Maximum frequency} \\
	    \noalign{\smallskip}\hline\noalign{\smallskip}
	  	  
	      & \SI{2}{mA}   & \multirow{6}{*}{$\left.\rule{0pt}{48pt}\right\}$\SI{110}{kHz}}   \\
		  & \SI{20}{mA}  & \\
		  & \SI{200}{mA} & \\
		  & \SI{2}{A}    & \\
		 \multirow{-5}{*}{\SI{2}{V}$\left.\rule{0pt}{40pt}\right\{$} & \SI{20}{A}    & \\
		  \SI{1}{V} $\lbrace$  & \SI{100}{A} & \\ 	 
		 
	\bottomrule
		\end{tabular}
	\label{tab:TAranges}
\end{table}

\begin{figure}[h]
    \centering
    \includegraphics[width=3.5 in]{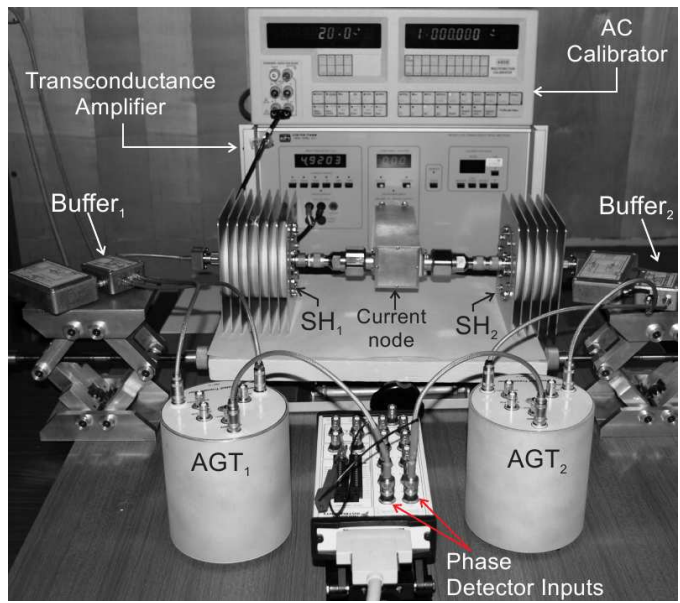}
    \caption{A photo of the experimental setup for phase comparison of high current shunts. The shunts shown are being entirely realized at INRIM, $\mathrm{SH_{1}}$ for currents up to \SI{20}{A} and $\mathrm{SH_{2}}$ for currents up to \SI{7}{A}. $\mathrm{SH_{1}}$ is made of metal film ($\cong\SI{50}{\mu m}$ thick) welded on special ceramic substrate; $\mathrm{SH_{2}}$ is made of thick film deposited on ceramic substrate by coating technology.} 
    \label{fig:PhotoSystem}
\end{figure}

A picture of the system supplying the current to the input of two shunts and the constituents of the phase comparator  is shown in figure \ref{fig:PhotoSystem}.

\subsection{The active guarded transformers}
Two AGTs are employed to transmit the voltage between the output of the shunts and the inputs of the phase detector. These transformers are of identical construction and each of them is made as a special type of double stage transformer. This structure was selected in order to increase the impedance of the transformer, especially at the intermediate frequencies. In fact, one of the core acts as the magnetizing element and the relevant winding is driven by a buffer. In this way, the primary winding only requires a negligible current sufficient to establish the correct balance between the input voltage and the electromotive force produced on the primary winding by the magnetizing core. So, the impedance, as seen from the output of the shunt, is high and there is only a negligible voltage drop on the cable due to this current. For wideband operation nanocrystalline cores were employed and all the windings were limited to 30 turns. The resulting operative bandwidth of the AGT is between \SI{100}{Hz} (at \SI{1}{Vrms}) and, with acceptable flatness (\SI{< 3}{dB}), to more than \SI{2}{MHz}, with a resonance peak at about \SI{5}{MHz}. However, over \SI{500}{kHz}, the phase shift becomes somewhat large (\SI{> 3}{mrad}) for high accuracy measurements. Each transformer has three separate screens, one for the magnetizing, one for the primary and the other one for the secondary winding. These screens can be connected to suitable voltages, in order to separate the influences of the capacitance between the two windings, thus increasing the common mode rejection.

\begin{figure}[h]
    \centering
    \includegraphics[width=3.5 in]{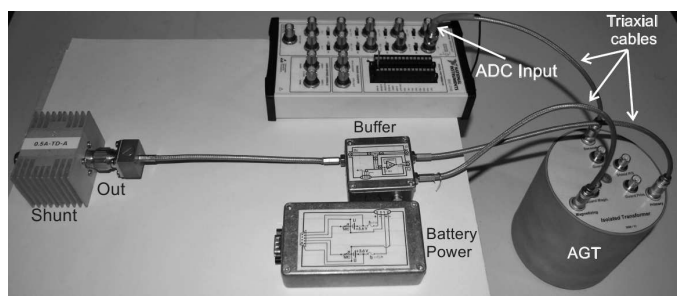}
    \caption{Particulars of the cables and the buffer for the magnetizing winding of the active guarded transformer.} 
    \label{fig:AGT_Buff}
\end{figure}

The voltages at the magnetizing windings are supplied by unity gain FET-input precision buffer, which has a flat bandwidth over \SI{10}{MHz}. Each buffer has low distortion of \SI{74}{dB} at \SI{5}{MHz}, output current of \SI{70}{mA} and input noise less than \SI{7}{nV/\sqrt{Hz}}. 

For frequencies below \SI{200}{Hz} the input impedance of the magnetizing winding of the AGTs becomes very low. In order to prevent damage of the buffer a current limiter, a resistor with nominal value of \SI{50}{\Omega}, is connected between the output of the buffer and the magnetizing winding of the AGTs.  
This resistance at the output of the buffer was then found to be a source of systematic error in the lower frequency range, where higher currents are required from magnetizing winding of the transformer. Each buffer is supplied by two independent \SI{1000}{mAh} Li-ion batteries, closed in a separate container, which can be connected to the buffer when this one is used. A proper screen, within the container is connected to the screens of the cable, the transformer and the one that surrounds each buffer, so that the entire contents of each AGT are enclosed in a screen that can be connected to a proper guard potential. A suitable battery charging device recharges these batteries from time to time.

\subsection{Asynchronous digital phase detector}
The first prototype of the digital phase detector was built by means of a commercial board containing four analog-to-digital converters (ADCs). The board is controlled by means of a personal computer through a peripheral component interconnect (PCI) bus. The independent input channels, each one with resolution of \SI{16}{bits}, can be configured to digitalize analog signals simultaneously at sampling rate up to \SI{1}{MSamples/s}. Coupling the two inputs of the digitizer permits the realization of a single unit digital phase detector. Recently, the arrangement used to realize  a wideband digital phase detector (WDPD) has been further extended in an experiment, which requires two distinct units of WDPD, for the characterization in terms of phase of a double electromagnetic device \cite{Trinchera11}.         
The starting condition for all inputs is given from a software selectable trigger. The inputs can be coupled in ac or dc and there are 8 voltage ranges from \SI{\pm 0.2}{V} to \SI{\pm 42}{V}, respectively. The on-board deep memory is \SI{64}{MB} and it is possible to switch on the on-board anti-aliasing filters by software control.
The phase relation of the two signals between the outputs of the active transformers and the inputs of the phase detector is maintained by connectors and cables with the same characteristics. Consecutive bursts of samples of the two inputs are acquired and processed to derive the phase difference between the two inputs as described in section \ref{sec:AcquisiComput}.

\section{Acquisition and computational algorithms}
\label{sec:AcquisiComput}

The analog voltages at the output of the shunts are simultaneously sampled and acquired by means of a program entirely developed in Matlab. The acquisition routine performs calls at a low level through the instrumentation control Toolbox directly to the driver of the digitizing system. In this way fast and immediate transfer of large quantities of data are shifted from virtual instrument and saved to removable devices. In the acquisition program there are two main routines. The initialization routine sets the initial timing parameters as well as the voltage level and the list of frequencies of the alternating calibrator, according to the parameters required by the transconductance amplifier, in order to supply the desired current at the input of the shunts under comparison. The parameters of the digitizing board, e.g. the vertical resolution and the sampling rate, are set by taking the depth of the internal on-board memory into account, in order to maximize the performances, and the specifications given by the asynchronous configuration.
    
The acquisition routine performs several steps. Starting from the frequency selected from the previous list, typically in the range from \SI{500}{Hz} to \SI{100}{kHz}, there are two processes namely time-depend  and acquire-transfer. The first one is composed of the main heating-time (not lower than \SI{15}{min} and depends on the shunts under comparison and the working current, the higher the current the higher is this time) and transition-time (time required to the calibrator to stabilize its output level when the frequency is changed). While the main heating-time depends on the shunts through the well known phenomena of Joule heating, the transition time, in the approximation of shunts with low losses, depends on the calibrator. Acquire-transfer process allows firstly the digitalization of the analog signal in a certain number of burst of nominally simultaneous samples. 
  
For each frequency the program acquires a selected number of bursts of nominally simultaneous samples. The number of samples per burst, chosen by the operator, should have sufficient periods for each burst and an adequate statistic to compute reliable measurement results and to indicate and exclude possible outlayers. 
The sample rate, for the best time definition, is generally chosen near the maximum value allowed by the board about \SI{1}{MSamples/s}; the output voltage of the shunts is digitized and the array of samples collected are shifted to a permanent data file on the hard disk. 
In the second part, the program analyses the burst of nominally simultaneous samples following partially a similar procedure as reported in \cite{Bosco11}.

In the samples collected from the first channel the transitions from negative to positive value are identified. A suitable number of samples is added after every transition to avoid counting the transitions due only to the noise. 
From the number of transitions and the sampling rate, a first tentative frequency of the signal is evaluated.
The mean value $\langle \mathrm{A_m} \rangle$ and the quadric mean (rms) of the fundamental of the signal $\mathrm{A_{rms}}$, which are supposed to be sinusoidal, are evaluated respectively by the relations:

\begin{eqnarray}
\label{eq:MeanValue}
\left \langle \mathrm{A_{m}} \right \rangle = \frac{1}{\mathrm{N_{f-l}}}\sum_{\mathrm{i=1}}^{\mathrm{N_{f-l}}} \nu_{\mathrm{i}} \\
\mathrm{A_{rms}} =\left( \frac{1}{\mathrm{N_{f-l}}}\sum_{\mathrm{i=1}}^{\mathrm{N_{f-l}}} \left(\nu_{\mathrm{i}} - \left \langle \mathrm{A_{m}} \right \rangle \right)^{2} \right)^{1 \over 2} 
\end{eqnarray}

where $\mathrm{N_{f-l}}$ is the number of samples between the first and the last transition.
The phases of the approximating sinusoid near the first and the last transition are then evaluated by the relation:

\begin{equation}
\label{eq:EqPhase}
\mathrm{\varphi_{i}} = \arcsin \left(\frac{\nu_{\mathrm{i}}-\left \langle \mathrm{A_{m}} \right \rangle} {\sqrt{2}\cdot \mathrm{A_{rms}}}\right)
\end{equation}

and these values, through a best fit approximation, are employed for a more accurate guess of the period.

The sine approximations of both bursts of the simultaneous samples are evaluated by means of the 4 parameters sine fit. The frequency is further adjusted and the sine and cosine components are computed. This process is iterative and it is repeated until the change of the frequency is less than 1 part in $10^9$.

Eventually the sine approximations of both bursts of the simultaneous samples are evaluated by means of the tree parameters sine fit. The phase difference between the two signals for each couple of simultaneous bursts of samples is then evaluated by:

\begin{equation}
\label{eq:PhaseDiff}
\mathrm{\Phi} = \arcsin \frac{\sin_{\mathrm{c_1}}}{\cos_{\mathrm{c_1}}}-\arcsin \frac{\sin_{\mathrm{c_2}}}{\cos_{\mathrm{c_2}}}
\end{equation} 

where the  $\mathrm{\sin_{c_{1,2}}}$ and $\mathrm{cos_{c_{1,2}}}$ are the sine and the cosine components, respectively, derived from the three parameters approximation of the two signals, and the phase is brought in the interval [0-2$\pi$] by a proper addition of 2$\pi$.

\section{Determination of phase difference between shunts}
\label{sec:DetPhasDiff}

The procedure adopted for the determination of the difference of the phase of the two shunts consists of two measurement sequences for every current and frequency, with different connections of the two shunts in the series. So, a complete measurement requires:

\begin{itemize}
\item The connection of the two inputs to the special current T-connector, which is carried out by the operator. One of the shunts is connected to the Low position, with the external screen connected at the ground potential. The other one is connected at the High position, with the external screen connected to the other end of the current generator.
\item The outputs of the shunts will be connected to the windings of the double stage transformers. The screens of the primary windings of the AGTs are put at the external side of the output of the shunts.
\item The buffers that supply the magnetizing windings of the active transformers are activated, after checking that their batteries are correctly charged.
\item The measurement program is started by the operator, who selects all the options: the current, the frequencies number of repetitions, the file where the data are stored, etc.
\item The program supplies the suitable current at the frequencies selected, acquires the samples and stores them into the computer.
\item The position of the shunts is then reversed on the series by connecting them in the opposite locations of the special T-connector, while the inputs of the two transformers and their screens are connected, with respect to the current supplying system, in the same position as before.
\item The measurement program is started again and the data for this connection of the shunts are acquired.
\end{itemize}

\subsection{Post-processing of the data}
\label{sec:DataProcess}

The first processing of the samples was performed, on the basis of the decision of the operator, immediately after the acquisition or in a successive time. In both cases the result of this processing is the difference between the phases of the two shunts under comparison in each single burst of measurements.
The results were then, after a visual inspection by the operator of the graphs for checking the consistency of the data collected, which is similar to the filtering process, summarized in the mean values and the standard deviations for both positions of the shunts.

Recently, a new algorithm was introduced: it performs a further analysis based on a post-processing filter immediately after the phase difference calculation. The results obtained from the algorithms are validated by direct comparison with the ones obtained from the visual operator inspection of the data.

\subsection{Determination of phase difference}
\label{sec:DiffPhaseDet}

\begin{figure}[h]
    \centering
    \includegraphics[width=3.5 in]{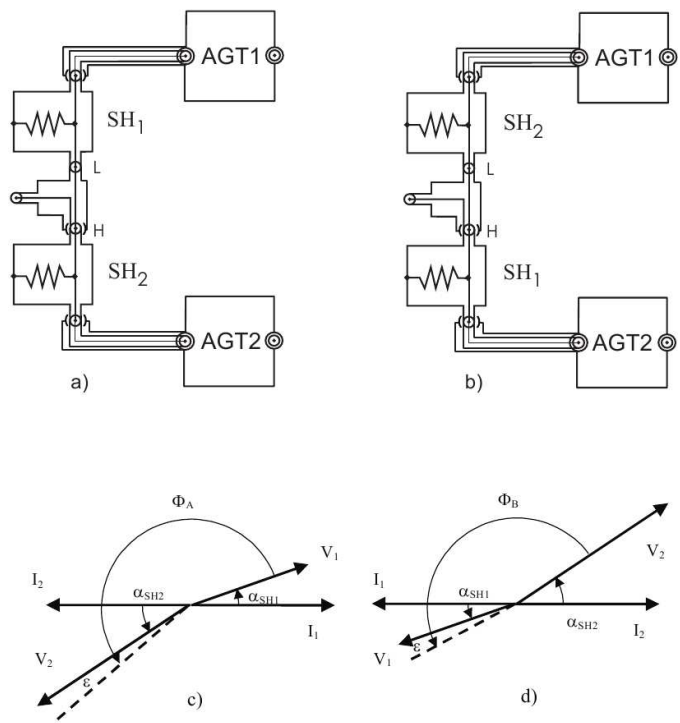}
    \caption{a) and b) Schemes of the two connections of the shunts in the measurement procedure. c) and d) Diagrams of the voltages and currents in the two connections a) and b) (the shunt phases $\alpha_{\mathrm{SH1}}$ and  $\alpha_{\mathrm{SH2}}$ and the systematic error $\epsilon$ are very exaggerated for explanation purpose).} 
    \label{fig:MeasSystem}
\end{figure}

The inversion of the shunts in the measurement procedure, represented in the schemes of figure \ref{fig:MeasSystem}(a) and (b), has been introduced for the compensation of the systematic errors in both the digitizers and the AGTs. In fact the structure of the transformers, of the digitizers and of the screens is the same, with presumably all the systematic errors while the signals at their input change by the inversion of the shunts.           

The voltages and the currents of the two conditions a) and b)  are shown in the phase diagrams of figure \ref{fig:MeasSystem}(c) and (d). In the series connection, while the input internal parts of the coaxial inputs of the shunts are connected together, the currents $I_1$ and $I_2$ are equal in amplitude but opposite to each other in phase. So, $\alpha_{\mathrm{SH_1}}$ and $\alpha_{\mathrm{SH_2}}$ are  the phase of both shunts, respectively. The measurement of the phase differences of the signals, $\Phi_A$ and $\Phi_B$ taken with the two different connections of the shunts are then related to the phase of the two shunts, $\alpha_{\mathrm{SH_1}}$ and $\alpha_{\mathrm{SH_2}}$, by the relations:

\begin{eqnarray}
\label{eq:diffPhase}
\left(\pi + \alpha_{\mathrm{SH_2}} \right) - \alpha_{\mathrm{SH_1}} = \Phi_{\mathrm{A}} - \epsilon \\
\left(\pi + \alpha_{\mathrm{SH_1}} \right) - \alpha_{\mathrm{SH_2}} = \Phi_{\mathrm{B}} - \epsilon 
\end{eqnarray}

So, by taking a combination of the values with the shunts in the reverse position, it is possible to cancel the effect of the phase error between the two channels $\epsilon$ and the phase difference is evaluated as:

\begin{equation}
\label{eq:EqdiffPhase}
\alpha_{\mathrm{SH_2}} - \alpha_{\mathrm{SH_1}} = \frac{\Phi_{\mathrm{A}} - \Phi_{\mathrm{B}}}{2}
\end{equation}	 

while the quantity:

\begin{equation}
\label{eq:epsilon}
\epsilon = \frac{\Phi_{\mathrm{A}} - \Phi_{\mathrm{B}}}{2} - \pi
\end{equation}	

is the phase difference between the two channels and it is specific for the phase comparator and its stability for a given frequency indicates the good functioning of the measurement system.
Eventually, the phase difference of two-shunt comparison is computed by equation (\ref{eq:EqPhase}) and its standard deviation is obtained by the square root of the quadratic composition of the components.

\section{Tests and results}

\subsection{Measurement performed on the elements of the phase comparator}

The phase comparator has been characterized for its functionality and for the repeatability of the results. In particular, several tests have been performed, by connecting its elements in different combinations, in order to determine the contribution of the elements at the phase-frequency characteristic of the comparator. 

The tests are performed in conditions far from experimental implementation. This leads to precise measurements of the phase difference of the digital phase comparator and active guarded transformer which are the key elements of the phase comparator.       

A first test was performed with the first prototype of the digital phase detector by connecting the two channels of the digitizing system, without the AGTs, in parallel to the same voltage input at the level of about 1 V with two short cables of almost the same length. After the acquisition, the samples were analysed by the algorithm for the determination of the phase difference, equation (\ref{eq:PhaseDiff}); the phase difference between the two channels of the phase detector was computed as a frequency function.
The results of the phase determination ($\mathrm{CH_{2}-CH_{1}}$) and the standard deviation of the values obtained, s, are reported in table \ref{tab:UncSteps}.

In the second test, in one of the two channels of phase detector the AGTs was inserted, and the difference in phase between this condition and the previous one is evaluated. This test was repeated for each AGT which was put in the two different channels (i.e., fixing $\mathrm{CH_{1}}$ as reference $\mathrm{AGT_{1}}$ and $\mathrm{AGT_{2}}$ are inserted in sequence to the channel $\mathrm{CH_{2}}$ of the phase detector; the measure is repeated by inverting the rule of the channels). The phase shift evaluated for each of the two AGTs is evaluated by means of the procedure described in section \ref{sec:DiffPhaseDet} and reported in table \ref{tab:UncSteps} as a function of the frequency with the mean value of the combined standard deviations.
In the third test the phase difference of the AGTs, evaluated by interchanging the two digitizing channels at the output of the two AGTs supplied in parallel, was derived.

In the last test, a special T-connector for inverting the Low and the High terminal was built and the two AGTs were associated to the two channels but the result was evaluated by two determinations and the phase difference is computed by means of equation (\ref{eq:EqdiffPhase}). The values of the differences between $\mathrm{AGT_1}$ and $\mathrm{AGT_2}$ and the results of this last measurement are almost equal, thus showing that different methods are congruent.          

\begin{table}[h]
	\caption{Results of the tests of the different elements of the phase comparator,where values at various frequencies are in (mrad), applying a constant voltage of \SI{1}{V} in the input.}
	\centering
		
	\begin{tabular}{c|c@{.}lc@{.}lc@{.}lc@{.}lc@{.}lc@{.}lc@{.}lc@{.}l} \toprule
	\textbf{}
	& \multicolumn{2}{c}{\SI{500}{Hz}}
	& \multicolumn{2}{c}{\SI{1}{kHz}}
	& \multicolumn{2}{c}{\SI{2}{kHz}}
	& \multicolumn{2}{c}{\SI{5}{kHz}}
	& \multicolumn{2}{c}{\SI{10}{kHz}}
	& \multicolumn{2}{c}{\SI{20}{kHz}}
	& \multicolumn{2}{c}{\SI{50}{kHz}}
	& \multicolumn{2}{c}{\SI{100}{kHz}} \\
	
		\midrule
		\small $\mathrm{(CH_2-CH_1)}$  & 0&003 & 0&006 & 0&013 & 0&034 & 0&069 & 0&140 & 0&352 & 0&712 \\ 
		\cmidrule(r){1-1}
		\small s                      & 0&001 & 0&001 & 0&001 & 0&001 & 0&001 & 0&001 & 0&003 & 0&006 \\
		\midrule \midrule
		\small $\mathrm{Buffer_1 \& AGT_1}$    & -0&036 & -0&006 & 0&003  & 0&012  & 0&025  & 0&054  & 0&156  & 0&404  \\  
		\cmidrule(r){1-1}
		\small $\mathrm{Buffer_2 \& AGT_2}$    & -0&037 & -0&006 & 0&002  & 0&011  & 0&024  & 0&055  & 0&141  & 0&341  \\
		\cmidrule(r){1-1}
		\small s          & 0&002 & 0&002 & 0&002 & 0&002 & 0&002 & 0&002 & 0&004 & 0&009 \\
		\midrule \midrule
 \small AGTs exchanged  & -0&001 & 0&000 & 0&000 & 0&000 & 0&000 & 0&001 & 0&010 & 0&054\\			    
		\bottomrule

	\end{tabular}
	\label{tab:UncSteps}
\end{table}

\subsection{Stability and phase-frequency dependence of the digitizer}
\label{sec:StabDigit}
The digitizer \cite{PCI5922} has been characterized in terms of long time stability applying in the inputs, $\mathrm{CH_{1}}$ and $\mathrm{CH_{2}}$, the same alternating voltage. A short twin-T coaxial cable equipped with three BNC connectors is built; the structure of the cable is made as symmetrical as possible. The measuring sequence and several parameters as stabilization and integration times, list of frequencies and voltage level, are identical to those adopted during normal use of the phase comparator (see section \ref{sec:AcquisiComput}). However, there are some slight differences compared to the procedures of post-processing described in section \ref{sec:DataProcess} and difference phase computation (section \ref{sec:DiffPhaseDet}), inasmuch, the contribution to the phase difference of the digitizer and cables is several times lower than the contribution of other elements of the comparator. This means that for the characterization of the digitizer in terms of phase it is not necessary to adopt the entirely measuring sequence as described in section \ref{sec:DetPhasDiff}. 

We are interested in exploring the behaviour, in terms of linearity and noise, of the phase difference between adjacent inputs of the digitizer system at various frequencies at a fixed sampling frequency. Single tone electric signals with well known electric parameters are suitable for this purpose. Timing constraints occurred, which became relevant for wideband digital phase detector because of the time quantization.

For this purpose two high speed low-distortion and noise waveform synthesizer are used. The first one is a commercial instrument\footnote{Tektronix AFG3252 dual channel arbitrary function generator} with vertical resolution of \SI{14}{bits} and update rate up to \SI{2}{GSamples/s}; the second one is a home made synthesizer \cite{Pogliano2011} with greater vertical resolution, \SI{16}{bits} and update rate up to \SI{1}{GSamples/s} per channel. Several tests should be employed with both synthesizers, but for phase test the choice fell on the commercial synthesizer, because it is able to change its sampling frequency maintaining  the number of points constant, which are fixed to 16384 samples from \SI{1}{kHz} to \SI{100}{kHz}, of the single tones synthesized. 

\begin{figure}[h]
    \centering
    \includegraphics[width=3.5 in]{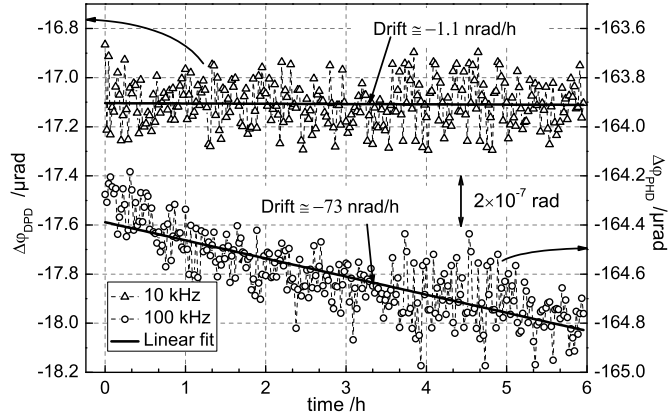}
    \caption{Drift of the digital phase detector setting the sampling frequency at \SI{10}{MSamples/s} and digitizer range to \SI{2}{V_{pp}}. For simplicity only measurements at \SI{10}{kHz} and \SI{100}{kHz} are reported.} 
    \label{fig:Drift_digit}
\end{figure}

The measuring sequence has been cycled for several hours and figure \ref{fig:Drift_digit} shows the results obtained after the computation and the filtering procedure. The time between cycles is about \SI{80}{s} and withal each cycle is composed by a fixed number of frequencies, typically \SI{8}; the stabilization time of the synthesizer is set to \SI{10}{s}. The range of both input channels of the digitizer is set to \SI{2}{Vpp} and they are coupled in dc, whereas the output impedance of the synthesizer is set to high-load; the sampling frequency is fixed to \SI{10}{MSamples/s}.    

\begin{figure}[h]
    \centering
    \includegraphics[width=3.5 in]{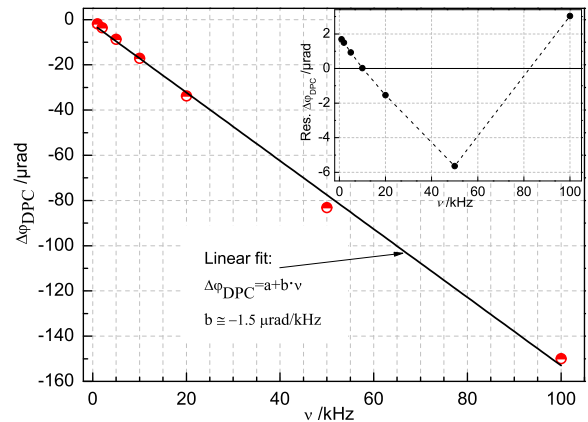}
    \caption{Behaviour of the phase difference of the digital phase detector at various frequencies.} 
    \label{fig:Ph_vs_freq}
\end{figure}

Figure \ref{fig:Ph_vs_freq} shows the behaviour of the phase difference of the digital phase detector, $\Delta\varphi_{\mathrm{DPC}}$, at various frequencies \footnote{The digitizer used is the PCI-5922, at \SI{10}{MHZ} of sampling frequency the vertical resolution is of \SI{18}{bits}(see \cite{PCI5922} for further detailed specifications). Characterization of the digitizer at lowest sampling frequency but highest vertical resolution, using a Programmable Josephson Voltage Standard (PJVS), is reported in \cite{Overney2011}. Further characterization of the same digitizer using a sampling frequency up to \SI{1}{MHz} is reported in \cite{Rietveld2011}.}.

\begin{figure}[h]
    \centering
    \includegraphics[width=3.5 in]{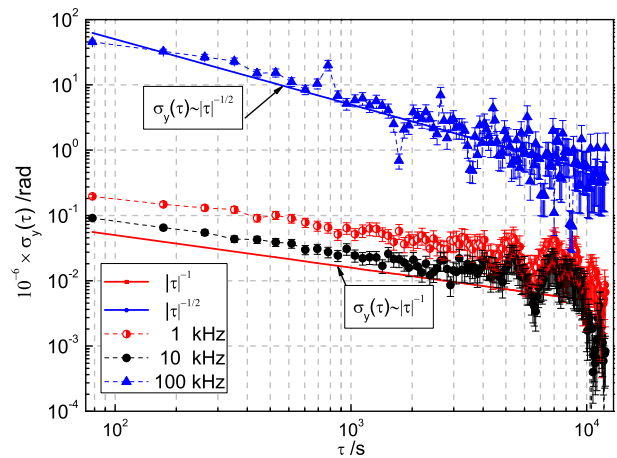}
    \caption{Allan deviation giving phase noise of the digital phase comparator (digitizer NI PCI-5922) applying  synthesized single tone alternating waveforms in its inputs.}   
    \label{fig:AllanDev}
\end{figure}

The intrinsic phase noise of the digital phase comparator is given in figure \ref{fig:AllanDev}. The measurements are performed at the same sampling frequency by applying single tone alternating waveforms with frequencies ranging between \SI{500}{Hz} and \SI{100}{kHz} and constant amplitude. 

Following \cite{IEEESTD1139}, the functional characteristic of independent noise process used in modelling phase and frequency instability of oscillators in the time domain is expressed as  $\sigma_{\mathrm(y)}(\tau) \propto \vert\tau\vert^{\mu/2}$.
 
The solid lines, see figure \ref{fig:AllanDev}, represent the phase instability of the digital phase detector modelled  in terms of independent noise process. For frequencies up to \SI{10}{kHz} the dependence of $\sigma_{\mathrm(y)}(\tau)$ is $\tau^{-1}$, which seems to indicate that the intrinsic noise contribution of phase comparator arises from the process due to flicker phase noise or white phase noise. Of course, modified Allan deviation, $\mathrm{Mod}~\sigma_{\mathrm(y)}(\tau)$, yields different dependence on $\tau$ in the presence of noise process mentioned above; we compute also the $\mathrm{Mod}~\sigma_{\mathrm(y)}(\tau)$ and the dependence was always $\tau^{-1}$; therefore it seems that the flicker phase noise describes well the phase fluctuation observed by means of long time measurements, when the phase detector is excited by alternating signals up to \SI{20}{kHz}. Always considering $\sigma_{\mathrm(y)}(\tau)$ the dependence on $\tau$, for frequency up to \SI{100}{kHz}, is $\tau^{-1/2}$. This leads to consider that the noise of the phase detector, observed for frequencies above \SI{50}{kHz}, may be given in terms of noise process namely white frequency modulation.

\subsection{Measurement performed on the shunts}

The phase comparator was employed for a comparison of existing shunts that are built by different European laboratories [14]. For currents of \SI{20}{A} and for frequencies up to 20 kHz, where the instability and the interference of the supplying current is not prevalent, the standard deviations of the phase differences measured were generally contained within \SI{10}{\mu rad}, but they raised to a mean value of about \SI{70}{\mu rad} at \SI{100}{kHz}. 
In the comparison a lot of data about the phase of different shunts taking part in the measurements were derived. At \SI{10}{A}, the expanded uncertainties of the phase differences were less than \SI{6}{\mu rad} up to \SI{10}{kHz} and less that \SI{40}{\mu rad} at \SI{100}{kHz}. The uncertainties increase for higher current, also due  to the effect of the variations of the phase between one current and the subsequent one.

\subsection{Contribution to uncertainty of the phase comparator}

Analysis of the measurements uncertainties was given in \cite{Bosco11}. In this paper we report only the identification and evaluation  of the contributions to the phase difference of the comparator with the enhanced version of digital phase detector. Some of them are evaluated by means of specific experiments; the assumptions made permit to evaluate single or composed effect to the phase difference. 
Table \ref{tab:UncBudget} lists the contributions arising from digital phase detector and constitutes of phase comparator. The meaning, the determination of those components are explained in \ref{appe:ExpUnc}.   
The largest contribution to the phase difference, at \SI{100}{kHz}, stems from active guarded transformer.

\begin{table}[h]
	\caption{Uncertainty budget of the digital phase detector and constituents of the phase comparator at \SI{10}{kHz} and \SI{100}{kHz}.}
	\centering
	\begin{threeparttable}[b]	
	
	\begin{tabular}{@{}lccc@{}} \toprule
        &\multicolumn{3}{c}{Uncertainty of the phase comparator in (\SI{}{\mu rad})} \\
	    \cmidrule(r){2-4}
	    &\multicolumn{3}{c}{SS1\tnote{1} @ \SI{10}{MHz} sampling frequency} \\
	    \cmidrule(r){2-4}
        Contribution to uncertainty                      & Type & \SI{10}{kHz} & \SI{100}{kHz}\\	    
	     \midrule
		\ Phase detector                             &   &      &      \\
		\cmidrule(r){1-1}
		\small Repeatability of 20 measurements             & A & \SI{0.10}{}            &  \SI{0.18}{}  \\
	    \small Clock jitter (see \cite{PCI5922})            & B & $\leq$\SI{0.18}{} & $\leq$\SI{1.8}{} \\
	    \small Drift within 3 hours\tnote{2}                & B & \SI{0.003}{}           & \SI{0.2}{}\\
	    \small Phase linearity within \SI{700}{mV}\tnote{3} & B & \SI{0.28}{}           & \SI{1.5}{} \\
	    \cmidrule(r){1-1}
	    \cmidrule(r){1-1}
	    \small $\mathrm{RSS_{Phase~detector}}$      &  & \SI{0.35}{} & \SI{2.3}{} \\
	    \midrule
		
		\ Comparator constituents                             &   &      & 		\\ 				
        \cmidrule(r){1-1}	    
	    \small $\mathrm{Buffer_1~\&~AGT_1}$\tnote{4}             & B & 2     & 9 \\
	    \small $\mathrm{Buffer_2~\&~AGT_2}$\tnote{4}             & B & 2     & 9 \\
	    \small Cable connections (see \ref{sec:UncCables})       & B & 0.02  & 0.3 \\
	    \cmidrule(r){1-1}
	    \cmidrule(r){1-1}
		\small $\mathrm{RSS_{Constituents}}$                      &   & 2.83 & 12.7 \\
		\midrule
		\small $\mathrm{RSS_{Tot}}$ \tnote{5}                    &   & 2.85 & 13 \\
		\bottomrule

	\end{tabular}
	\begin{tablenotes}
    \item[1]\footnotesize NI PCI-5922: flexible resolution digitizer, \SI{24}{bit} resolution up to \SI{500}{kSamples/s}, ranging to \SI{16}{bits} at \SI{15}{MSamples/s}.
    \item[2]\footnotesize A complete measure of phase difference between two shunts is made always within \SI{3}{hours} (see figure \ref{fig:Drift_digit}).
    \item[3]\footnotesize Typically, for shunts with unlike nominal values the phase detector is excited at different voltages.  A voltage difference, $\Delta V~\cong~0.7V$, will produce an uncertainty given by $\Delta V\frac{\partial\varphi}{\partial V}$, i.e., at \SI{10}{kHz} is \SI{0.28}{\mu rad} and at \SI{100}{kHz} is \SI{1.5}{\mu rad}, respectively.
    \item[4]\footnotesize Phase difference of blocks made by buffer and active guard transformer.
    \item[5]\footnotesize $\mathrm{RSS_{Tot}=\sqrt{RSS^2_{Phase~Detector} + RSS^2_{Constituents}}}$.
  \end{tablenotes}
\end{threeparttable}
	\label{tab:UncBudget}
\end{table}

\section{Conclusions}
A wideband phase comparator for high current shunts is built at INRIM. The comparator has been  extensively used during the European project and is still used in the metrological chain of phase angle maintained by precise shunts. Several tests are performed and the repeatability of the measurements is in good agreement within expanded uncertainty (k=2), which for a group of shunts \SI{10}{A} is \SI{50}{\mu rad} at \SI{100}{kHz}. 

Tests performed on different elements of the comparator demonstrated that, although the constitutive elements (digital phase detector, AGTs, buffers and triaxial cables) have a not negligible phase shift , their stability and the compensation of the procedure improve the accuracy of the phase determination of at least an order of magnitude. By means of the comparator a set of shunts has been compared and the values obtained, which are verified by the coherence principle of closed loop schemes, show that they are accurate, stable and repeatable.
An enhanced version of this comparator is designed, which employs a new phase detector constructed around a faster digitizer board and improved circuit for annulling the current at the input of the AGTs. In this way, it would be possible to improve the repeatability of the measurements and to increase the operative range of the comparator. The contribution to extended uncertainty due to the faster phase detector is improved several time and it is estimated to be lower than \SI{0.5}{\mu rad} at \SI{10}{kHz} and \SI{3}{\mu rad} at \SI{100}{kHz} with (k=1).          

\section{Acknowledgments}

The authors wish to thank G. C. Bosco (INRIM) for his technical support during the characterization of first prototype of the phase comparator and F. Francone (INRIM) for the realization of adapters and triaxial cables.

\appendix

\section{Details on evaluation methodology for uncertainty components}
\label{appe:ExpUnc}

In this appendix we report a brief explanation about the different methodologies and joint test regarding the evaluation of uncertainty components that appear in table \ref{tab:UncBudget}.

\subsection{Wideband digitizer system}
\label{sec:UncDigit}
Input channels of the digitizer board should be imagined as black-boxes, and test separately analog and digital circuits is impossible. A powerful well known approach is to apply known electric signal, depending on the specific test that we would like to run. Here we are interested in evaluating the difference in term of phase between the two inputs of the digitizer board. By applying the same signal the phase difference at different frequencies, according to \ref{eq:EqPhase}, is computed (namely repeatability in table \ref{tab:UncSteps}. Other contributions are extrapolated from \cite{PCI5922} and computed from $\Delta\varphi=2 \pi \nu \Delta t$.

Two additional tests are performed in order to evaluate possible contributions to uncertainty of phase detector. The first one is related to long time instabilities described in section \ref{sec:StabDigit}.

\begin{figure}[h]
    \centering
    \includegraphics[width=3.5 in]{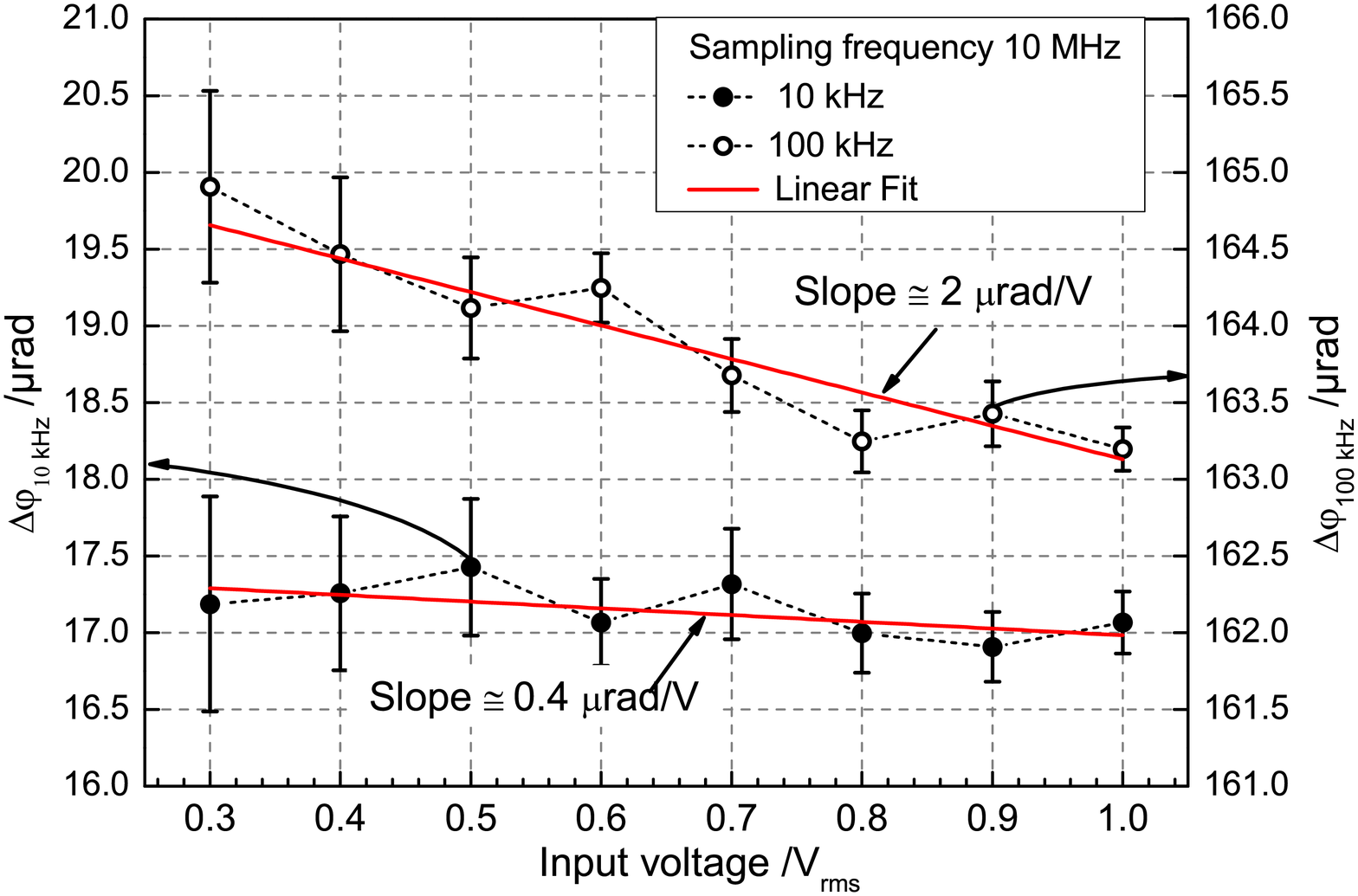}
    \caption{Behaviour of phase difference of digital phase detector at \SI{10}{kHz} and \SI{100}{kHz} varying the amplitude of the alternating signal. The sampling frequency of the digitizer is fixed to\SI{10}{MSamples/s} and the vertical range is set to \SI{3}{V}.} 
    \label{fig:PhaseVSAmplitude}
\end{figure}

The method of calculation, adopted to assign the phase of shunts \cite{Bosco11}, also requires the knowledge of the phase difference between the shunts having unlike nominal values. It is clear that the shunts with low nominal value in several comparison work far from nominal current, thus producing a lower output voltage. In a condition like this, the signals coming from the output of the shunts are different and in some comparisons a voltage difference higher than \SI{100}{mV} is reached. The dependence of phase difference, detected by means of a phase detector on the voltage applied is evaluated. The alternating voltage applied is ranging from \SI{300}{mV} and \SI{1}{V}. The test is performed at fixed frequencies (i.e., \SI{10}{kHz} and \SI{100}{kHz}) and the results are reported in figure \ref{fig:PhaseVSAmplitude}. The differences of phase between \SI{0.3}{V} and \SI{1}{V} obtained from the linear fit are interpreted as uncertainty due to phase-amplitude dependency.

\subsection{Uncertainty of connecting cables}
\label{sec:UncCables}
Characteristic impedance, $Z_{\mathrm{0}}=\sqrt{L/C}$, and propagation delay, $\Delta t=\sqrt{LC}$, are the most important characteristic of a transmission line, where $L$ and $C$ are the inductance and capacitance distributed along the line. A short coaxial cable as the RG-58, which is widely used in the field of low frequency metrology, of \SI{30}{cm} has $L\simeq\SI{76.8}{nH}$ and $C\simeq\SI{31.2}{pF}$; computing characteristic parameter one obtain $Z_{\mathrm{0}}\simeq\SI{49.6}{\Omega}$ and $\Delta t\simeq\SI{1.55}{ns}$. The propagation velocity is about \SI{1.966e+8}{m/s} or \SI{66.6}{\%} of the speed's light. 

The phase difference is computed starting from the definition of angular velocity $\omega=2\pi\nu=\Delta\varphi/\Delta t$; for the coaxial cable mentioned above the phase difference between input and output is $\Delta\varphi_{\SI{100}{kHz}}\simeq\SI{1}{mrad}$. Keeping in mind that for an ideal line the delay increases linearity with the length and the phase difference between its input and output will increase with the frequency, one should be able to compute the phase difference between two coaxial cables which differ slightly in terms of length. So, if their difference is about \SI{1}{mm} (e.g., the coaxial cables employed to build the twin-T cable, which is used to characterize the digitizer) the phase difference at their outputs, when the inputs are connected together. is $\Delta\varphi_{\SI{100}{kHz}}\simeq\SI{3.2}{\mu rad}$, respectively.

However, considering the cables used in the set-up phase comparator we can apply the same concepts to calculate the phase difference between the two legs, which is classified as a systematic error and can be eliminated by proper inversion procedure. This inversion procedure is not the same as the one used during the phase difference measure between shunts.

The effects of cable to phase difference is measured directly by the phase comparator. For this purpose the cables are inserted between the AC calibrator and the inputs of the phase detector, without inserting the active guarded transformers. We find that experimental data, at \SI{100}{kHz}, deviate from the theoretical prediction of about \SI{50}{\%}. The measurements are performed several times and the standard deviation are treated as uncertainty contribution inserted in table \ref{tab:UncBudget}.        

\subsection{Uncertainty of buffers}
The phase difference introduced by the buffers is evaluated by measuring procedure. Special adapters( triaxial to coaxial BNC) have been constructed in order to connect the output of the box buffer to the input of the digitizers (i.e., the AGTs are left out during this test). A substitution procedure has been adopted and the phase difference between the buffers,$\Delta\varphi_\mathrm{Buf1-Buf2}$, is computed starting from the single phase measurements of each buffer, $\varphi_\mathrm{Buf1}$ and $\varphi_\mathrm{Buf1}$. 

It has been found that the uncertainty due to the buffers is lower than \SI{3}{\mu rad} at \SI{100}{kHz}. This uncertainty is not reflected directly on the uncertainty of the AGT, which, however has already been included implicitly in the table \ref{tab:UncBudget}.     

\restoreapp  

\section*{References}
\bibliographystyle{unsrt}
\bibliography{MetrologiaPhaseComp}

\end{document}